\documentclass[iop]{emulateapj}

\usepackage{epstopdf}

\newcommand{\psizone}{1.15}

\slugcomment{Accepted for publication by ApJ Letters}

\shorttitle{Magnetic Braking Formulation}
\shortauthors{Matt et al.}

\begin{document}













\title{Magnetic Braking Formulation for Sun-Like Stars: \\Dependence on Dipole Field Strength and Rotation Rate}

\author{Sean P. Matt$^{1}$,
        Keith B. MacGregor$^{2}$,
        Marc H. Pinsonneault$^{3}$, and
        Thomas P. Greene$^{4}$}

\affil{$^1$Laboratoire AIM Paris-Saclay, CEA/Irfu Universit\'e
 Paris-Diderot CNRS/INSU, 91191 Gif-sur-Yvette, 
France; sean.matt@cea.fr}

\affil{$^2$HAO NCAR; kmac@ucar.edu}

\affil{$^3$Ohio State University; pinsonneault.1@osu.edu}

\affil{$^4$NASA Ames Research Center, M.S. 245-6, Moffett Field, CA
 94035-1000, USA; thomas.p.greene@nasa.gov}


\begin{abstract}

  We use two-dimensional axisymmetric magnetohydrodynamic simulations
  to compute steady-state solutions for solar-like stellar winds from
  rotating stars with dipolar magnetic fields.  Our parameter study
  includes 50 simulations covering a wide range of relative magnetic
  field strengths and rotation rates, extending from the slow- and
  approaching the fast-magnetic-rotator regimes.  Using the
  simulations to compute the angular momentum loss, we derive a
  semi-analytic formulation for the external torque on the star that
  fits all of the simulations to a precision of a few percents.  This
  formula provides a simple method for computing the magnetic braking
  of sun-like stars due to magnetized stellar winds, which properly
  includes the dependence on the strength of the magnetic field, mass
  loss rate, stellar radius, suface gravity, and spin rate and which
  is valid for both slow and fast rotators.

\end{abstract}


\section{Introduction} \label{sec_introduction}

The evolution of the rotational properties of sun-like stars exhibit
numerous trends that still lack accepted, quantitative explanation.
For example, during the pre-main-sequence phase, there is not yet a
comprehensive picture that explains the observed wide range of
rotation rates, nor the fact that a large fraction of stars rotate
relatively slowly \citep[e.g.,][]{herbstea07, irwinbouvier09,
  mattpudritz05, mattpudritz05l}.  During the main sequence phase, we
are still trying to understand, for example: the structure and
existence of multiple ``sequences'' apparent in the
rotation-period-versus-mass diagrams of young star clusters
\citep[e.g.,][]{Barnes:2003p1513, irwinbouvier09, meibomea11}; the
apparent ``saturation'' of angular momentum loss in fast rotators
\citep{MacGregor:1991p4316}; the overall, secular spin-down of stars
\citep{Kraft:1967p4433, Skumanich:1972p4350, Soderblom:1983p4391}; the
possibility of using rotational properties to measure stellar ages
\citep[``gyrochronology'';][]{Barnes:2003p1512, Barnes:2010p1157,
  Mamajek:2008p4506, meibomea11, Epstein:2012p4186}; and the
correlation of stellar activity with rotation, as well as the
``saturation'' of this activity in fast rotators
\citep[e.g.,][]{Saar:1999p3589, Pizzolato:2003p3630,
  Reiners:2009p3402, Wright:2011p4333}.

We know that magnetized stellar winds are important for extracting
angular momentum from stars during the main sequence \citep{parker58,
  Schatzman:1962p4464, Weber:1967p3752, Mestel:1968p4497} and likely
during pre-main-sequence \citep{Hartmann:1982p2888,
  Hartmann:1989p2909,mattpudritz05l}.  Thus, a prescription for
calculating the stellar wind torque as a function of stellar
parameters is a crucial ingredient in models for the rotational
evolution of stars \citep[e.g.,][]{Bouvier:1997p4099, bouvier08,
  denissenkovea10,Matt:2012p3910}.

Reliably computing the stellar wind torque requires knowledge of the
wind acceleration profile and the magnetic field geometry above the
surface of the star \citep[e.g.,][]{Mestel:1984p2919}.  Until a few
years ago, the only formulations available for computing stellar wind
torques \citep[e.g.,][]{Kawaler:1988p1012} were based upon analytic or
semi-analytic calculations that necessarily relied upon several
simplifying assumptions, such as that of spherical symmetry and the a
priori specification of the magnetic geometry, flow acceleration
profile, or both.  However, in real winds, all of the flow properties
are determined by a complex interaction between the wind driving
physics, stellar rotation, and dynamical interaction between the wind
and magnetic field, and these significantly deviate from spherical
symmetry.  Multi-dimensional, magnetohydrodynamical (MHD) simulations
provide a reliable method for computing dynamically self-consistent
wind solutions, but large parameter studies are needed, in order to
quantify how the physics scale with various parameters.  Thus,
although there exists a large body of literature on the subject,
reliably calculating stellar wind torques for a range of stellar and
wind properties, and in a way that is useful for stellar evolution
calculations, remains a challenging problem.

Using simulations of simplified, solar-like winds, \citet[][hereafter
MP08]{mattpudritz08II} carried out a small parameter study to
determine the dependence of the wind torque on the strength of the
magnetic field and mass outflow rate in the wind\footnote{Although the
  results were presented in the context of pre-main-sequence angular
  momentum loss, the simulations of MP08 are also appropriate for main
  sequence stars.}.  Their formulation for the torque (equation 15 of
MP08) differed significantly from previous formulations, notably in
the value of the exponent in the power-law dependence of the torque on
stellar properties, such as magnetic field strengh, mass loss rate,
and stellar radius.  This study adopted many of the same assumptions
as previous analytic work---ideal MHD, a rotation-axis-aligned dipolar
magnetic field, solid body stellar rotation, and spherically symmetric
thermodynamic properties at the stellar surface---but the simulations
did not require any assumptions about the kinematics of the flow nor
how the magnetic geometry was modified by the flow.  Thus, the MP08
torque is the most dynamically self-consistent formulation for the
torque from sun-like stars to date, and the implications for stellar
evolution are still being explored.

At the same time, this formulation is derived from simulations with
variations only in the magnetic field strength (relative to the mass
loss rate and surface gravity).  It does not fully capture the effects
of different rotation rates, different thermodynamic (or energetic)
properties of the wind, nor more complex magnetic geometries.  Given
the importance of computing stellar wind torques for a range of
stellar ages (and thus a range of rotation rates), the natural next
step is to extend the parameter study of MP08 to include variations in
both magnetic field strength and stellar rotation rate.  In this
letter, we present such a parameter study and, from these results,
derive the most generally applicable stellar wind torque formula to
date.

\section{Simulation Method and Parameter Study} \label{sec_method}

We use the simulation code and method described in MP08 to compute
steady-state wind solutions for sun-like stars with a dipole magnetic
field.  We briefly describe the method here, and the reader will find
further details in MP08 and references therein.  The code solves the
equations of ideal MHD under the assumption of axisymmetry and a
polytropic equation of state ($P \propto \rho^\gamma$).  In each
simulation, the numerical grid is initialized with a spherically
symmetric, thermally-driven Parker wind solution \citep{parker58},
plus analytic dipole magnetic field.  Once the simulations begin, the
wind solution relaxes to a steady state resulting from a dynamical
balance between the accelerating wind and rotating magnetic field.
The steady-state solution is entirely determined by the conditions
that are present at the base of the wind (the ``surface'' of the
star).

For a given initial magnetic geometry, unique wind solutions are
determined by dimensionless parameters, which can be given as three
velocity ratios---$v_{\rm A}/v_{\rm esc}$, $c_{\rm s}/v_{\rm esc}$,
and $f$ specified at the surface and equator of the star---plus the
adiabatic index $\gamma$.  Here, $v_{\rm A}$ is the Alfv\'en speed,
$v_{\rm esc}$ the gravitational escape speed, $c_{\rm s}$ the thermal
sound speed, and $f$ is the equatorial rotation speed divided by the
breakup speed,
\begin{eqnarray}
\label{eq_vavesc}
v_{\rm A}/v_{\rm esc} = B_* ({4 \pi \rho_*})^{-1/2}  (2 G M_* / R_*)^{-1/2}, \\
\label{eq_f}
f \equiv \Omega_* R_*^{3/2} (G M_*)^{-1/2},
\end{eqnarray}
where $B_*$ is the magnetic field strength at the stellar equator,
$\rho_*$ the mass density at the base of the wind, $G$ Newton's
gravitational constant, $M_*$ the stellar mass, $R_*$ the stellar
radius, and $\Omega_*$ the (solid body) angular rotation rate of the
stellar surface ($= 2 \pi / P_*$, where $P_*$ is the rotation period).

The torque formulation of MP08 is based upon 9 simulations with
variations in the parameter $v_{\rm A}/v_{\rm esc}$.  They also
presented 5 simulations with variations in the other parameters, $f$,
$c_{\rm s}/v_{\rm esc}$, or $\gamma$, which demonstrated that these
parameters affect the torque in a way that is not captured by their
fit formulation.  Motivated by the fact that the study of MP08 is
based on a relatively small number of simulations and by the
importance of precisely determining the torque for a range of stellar
rotation rates, the present work extends the parameter study of MP08
to include 50 simulations sampling a large range in both $v_{\rm
  A}/v_{\rm esc}$ and $f$.  For all simulations presented here, we
adopt a dipole magnetic geometry and fix $\gamma = 1.05$ and $c_{\rm
  s}/v_{\rm esc} = 0.222$, as in the ``fiducial'' case of MP08.



\begin{deluxetable}{cllrccc}
\tablewidth{0pt}
\tablecaption{Simulation Parameters and Results \label{tab_parms}}
\tablehead{
\colhead{Case} &
\colhead{$f$} &
\colhead{$v_{\rm A} / v_{\rm esc}$} &
\colhead{$\Upsilon$} &
\colhead{$r_{\rm A} / R_*$} 
}

\startdata

1 & 0.0000995 & 0.0753 & 27.5 & 5.15  \\
2 & 0.0000995 & 0.301 & 446 & 9.42  \\
3 & 0.0000995 & 1.51 & 13500 & 19.5  \\
4 & 0.000997 & 0.0301 & 4.11 & 3.31  \\
5 & 0.000997 & 0.0753 & 25.8 & 5.04  \\
6 & 0.000997 & 0.301 & 437 & 9.58  \\
7 & 0.000997 & 1.51 & 13000 & 19.4  \\
8 & 0.00393 & 0.0753 & 25.8 & 5.03  \\
9$^a$ & 0.00386 & 0.209 & 215 & 8.36  \\
10 & 0.00393 & 0.301 & 436 & 9.57  \\
11 & 0.00386 & 0.418 & 837 & 10.9  \\
12 & 0.00393 & 0.953 & 4900 & 15.8  \\
13 & 0.00393 & 1.51 & 13000 & 19.4  \\
14 & 0.0101 & 0.0753 & 25.6 & 5.01  \\
15 & 0.0101 & 0.301 & 432 & 9.50  \\
16 & 0.0101 & 1.51 & 12700 & 19.4  \\
17 & 0.0202 & 0.0753 & 24.9 & 4.95  \\
18 & 0.0202 & 0.301 & 417 & 9.31  \\
19 & 0.0202 & 1.51 & 11900 & 19.2  \\
20 & 0.0299 & 0.0753 & 24.0 & 4.86  \\
21 & 0.0299 & 0.301 & 395 & 9.04  \\
22 & 0.0299 & 1.51 & 10900 & 18.7  \\
23 & 0.0403 & 0.0753 & 22.8 & 4.74  \\
24 & 0.0403 & 0.301 & 367 & 8.73  \\
25 & 0.0403 & 1.51 & 9840 & 18.1  \\
26 & 0.0493 & 0.0753 & 21.6 & 4.62  \\
27$^a$ & 0.0493 & 0.235 & 212 & 7.66  \\
28 & 0.0493 & 0.301 & 341 & 8.44  \\
29 & 0.0493 & 1.51 & 8960 & 17.5  \\
30 & 0.0594 & 0.0753 & 20.1 & 4.49  \\
31 & 0.0594 & 0.301 & 312 & 8.12  \\
32 & 0.0594 & 1.51 & 8020 & 16.8  \\
33 & 0.0986 & 0.0301 & 2.70 & 2.82  \\
34 & 0.0986 & 0.0753 & 14.6 & 3.96  \\
35$^a$ & 0.0986 & 0.209 & 104 & 5.98  \\
36$^a$ & 0.0987 & 0.254 & 150 & 6.44  \\
37$^a$ & 0.0986 & 0.301 & 211 & 6.97  \\
38$^a$ & 0.0986 & 0.363 & 295 & 7.53  \\
39$^a$ & 0.0990 & 0.502 & 580 & 8.77  \\
40$^a$ & 0.0986 & 0.602 & 857 & 9.58  \\
41$^a$ & 0.0986 & 0.940 & 2130 & 11.8  \\
42$^a$ & 0.0986 & 2.10 & 10400 & 16.8  \\
43$^a$ & 0.0986 & 3.01 & 20500 & 19.3  \\
44 & 0.197 & 0.0753 & 5.39 & 2.83  \\
45 & 0.197 & 0.301 & 75.9 & 4.93  \\
46$^a$ & 0.197 & 0.495 & 212 & 6.26  \\
47 & 0.197 & 1.51 & 2120 & 10.5  \\
48 & 0.403 & 0.753 & 88.0 & 4.61  \\
49 & 0.403 & 1.51 & 370 & 6.30  \\
50 & 0.403 & 3.01 & 1470 & 8.33

\enddata

\tablenotetext{a}{Same Case used in the study of MP08.}

\end{deluxetable}

Table \ref{tab_parms} lists the input parameters $f$ and $v_{\rm
  A}/v_{\rm esc}$ for all 50 cases of the present study.  Twelve of
the simulations, as indicated in the table, are identical to those
used in MP08.

\section{Simulation Results} \label{sec_results}

From the steady-state wind simulations, we compute the total mass loss
rate ($\dot M_{\rm w}$) and the torque ($\tau_{\rm w}$) directly, by
integrating the mass and angular momentum flux over a closed surface
containing the star.  Table \ref{tab_parms} lists the value of
\begin{eqnarray}
\label{eq_upsilon}
\Upsilon \equiv {B_*^2 R_*^2 \over \dot M_{\rm w} v_{\rm esc}},
\end{eqnarray}
for each case.  The quantity $\Upsilon$ is a dimensionless way of
expressing the (inverse) mass loss rate resulting from the
simulations, it is physically related to the simulation parameter
$(v_{\rm A}/v_{\rm esc})^2$, and it is similar to the ``magnetic
confinement parameter'' of \citet[][\citealp{uddoula3ea08,
  uddoula3ea09}]{UdDoula:2002p984}.  In the following analysis, we
treat $\Upsilon$ as an independent variable of the simulations, even
though it is a {\it result} of the simulations.  We prefer to work
with $\Upsilon$, as it contains the more observable/predictable
quantity $\dot M_{\rm w}$, as opposed to $v_{\rm A}/v_{\rm esc}$, for
which one must specify the density at the base of the corona $\rho_*$
(see eq.\ [\ref{eq_vavesc}]).  Furthermore, for a given value of
$\rho_*$, the mass loss rate depends sensitively upon the values of
$c_{\rm s}/v_{\rm esc}$ and $\gamma$ (see MP08), as well as the
particular choice of boundary conditions, which we hold fixed in all
simulations.  Thus, the consideration of $\Upsilon$ as a controlling
parameter (instead of $v_{\rm A}/v_{\rm esc}$) significantly mitigates
the influence of $c_{\rm s}/v_{\rm esc}$, $\gamma$, and boundary
conditions choices in the resulting torque formulation.

\begin{figure}
\epsscale{\psizone}
\plotone{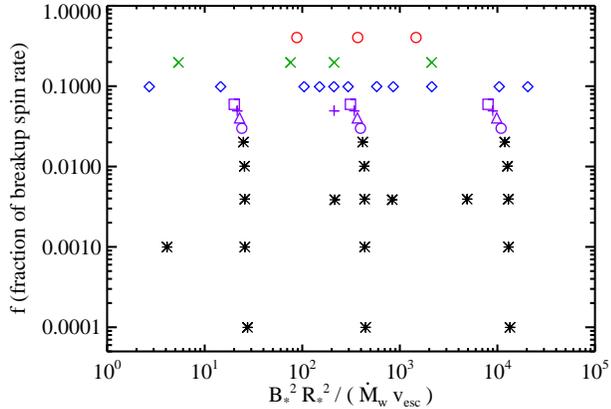}
\caption{Parameter space explored.  Each point represents a single
stellar wind simulation.  Symbols are as follows: asterisks
represent cases with $f \la 0.02$, circles are for $f \approx 0.03$
and 0.4, triangles are for $f\approx 0.04$, plusses are for
$f\approx 0.05$, squares are for $f\approx 0.06$, diamonds are for
$f\approx 0.1$, and X's are for $f\approx 0.2$.}
\label{fig_parmspace}
\end{figure}

Figure \ref{fig_parmspace} illustrates the range of parameter space,
in terms of $f$ and $\Upsilon$, spanned by all 50 simulations.  Each
symbol in the plot corresponds to a simulation listed in Table
\ref{tab_parms}, and the different symbol styles highlight different
ranges of spin rates.  The asterisks are for cases in the slow rotator
regime, where the rotation has a negligible influence on the wind
dynamics.  The remaining symbols represent cases in which rotation
affects the speed and collimation, and consequently the efficiency of
angular momentum loss, in the flow.  For reference, the appropriate
values for the solar wind are $f \approx 0.004$ and $\Upsilon$ within
a possible range of $\sim 10^{2-3}$ (e.g., for $\dot M_{\rm w} \approx
2 \times 10^{-14} M_\odot$ yr$^{-1}$ and an equatorial dipole field
strength of 1--5 Gauss).

In order to express the resulting torques in a useful and general way,
we consider the following.  In a steady-state wind, under the
assumptions of ideal MHD, the specific angular momentum extracted from
the star is equal to $\Omega_* r_{\rm A}^2$
\citep[e.g.,][]{Weber:1967p3752}, where $r_{\rm A}$ is the ``Alfv\'en
radius,'' the cylindrical radial location where the wind velocity
equals the local Alvf\'en speed.  In a three-dimensional flow, the net
torque on the star can be written
\begin{eqnarray}
\label{eq_tauw}
\tau_{\rm w} = \dot M_{\rm w} \Omega_* r_{\rm A}^2,
\end{eqnarray} 
where $r_{\rm A}^2$ is an average value \citep{Washimi:1993p3825}.
From the simulations, we compute $\tau_{\rm w}$ and $\dot M_{\rm w}$
directly, so we use equation (\ref{eq_tauw}) to define the
dimensionless Alfv\'en radius as
\begin{eqnarray}
\label{eq_rarstar}
{r_{\rm A} \over R_*} \equiv 
      \left({\tau_{\rm w} \over \dot M_{\rm w} \Omega_* R_*^2}\right)^{1/2},
\end{eqnarray}
which thus represents the square root of the dimensionless torque and
is unique for a given set of dimensionless simulation parameters.
Table \ref{tab_parms} lists the resulting values of $r_{\rm A} / R_*$
for each case.



\section{New Torque Formulation} \label{sec_formula}

\begin{figure}
\epsscale{\psizone}
\plotone{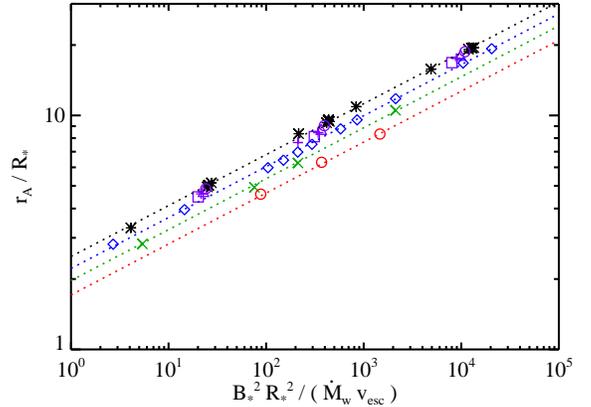}
\caption{Alfv\'en radius (eq.\ [\ref{eq_rarstar}]) versus $\Upsilon$.
 The symbols have the same meaning as in Figure \ref{fig_parmspace}.
 Dotted lines show the fit function (eq.\ [\ref{eq_fitra}] and
 [\ref{eq_fitvals}]) for $f = 0.001$, 0.1, 0.2, and 0.4, from top to
 bottom, which approximately correspond to a few of the rotation
 rates used in the parameter study.}
\label{fig_raplot}
\end{figure}

Figure \ref{fig_raplot} shows the value of the dimensionless Alfv\'en
radius versus $\Upsilon$ for the entire parameter study.  The study of
MP08 found the dimensionless Alfv\'en radius was well-represented by a
single power-law in $\Upsilon$, for all cases with the same spin rate.
In other words, 9 simulations in MP08 were fit by $r_{\rm A}/R_* = K
\Upsilon^m$, where $K$ and $m$ were dimensionless fit parameters.  In
Figure \ref{fig_raplot}, it is evident that the effect of rotation is
to modify the value of $K$, but not to significantly affect the
exponent $m$ (represented by the slope in the log-log plot).  Thus, we
are able to fit the 50 simulations here, with variations in both
$\Upsilon$ and $f$, with only one additional free parameter.
Specifically, we find that all of the data are well fit by the
function
\begin{eqnarray}
\label{eq_fitra}
%
%
{r_{\rm A} \over R_*} = K_1 \left[{\Upsilon \over (K_2^2 + 0.5 f^2)^{1/2}}\right]^{m}
\end{eqnarray}
where $K_1$, $K_2$, and $m$ are dimensionless fit constants.
The best fit values give:
\begin{eqnarray}
\label{eq_fitvals}
K_1 = 1.30,    \;\;\;\;\;\;\;
K_2 = 0.0506,    \;\;\;\;\;\;\;
m = 0.2177
\end{eqnarray}
These values differ only by a few percent from the equivalent fit
parameters of MP08 ($m$ and $K$) but should be considered to be more
precise here, due to the larger number of simulations fit.  Also, the
functional form of equation (\ref{eq_fitra}) simultaneously quantifies
the effect of the wind magnetization ($\Upsilon$) and the spin rate
($f$) on the effective Alfv\'en radius.  The dotted lines in Figure
\ref{fig_raplot} show equation (\ref{eq_fitra}) for four different
values of $f=$ 0.001, 0.1, 0.2, and 0.4 and using the best fit values
(\ref{eq_fitvals}).  Each dotted line goes through the symbols that
represent simulations with corresponding spin rates, illustrating how
well equation (\ref{eq_fitra}) fits the simulation results.

One can understand the functional form of equation (\ref{eq_fitra}) as
follows.  The most important factor for determining the Alfv\'en
radius is the strength of the magnetic field compared to the inertia
in the flow, $\Upsilon$.  The MHD simulations self-consistently
capture how the wind accelerates, how the magnetic field strength
varies with distance from the star, and how much of the total surface
magnetic flux will participate in the wind (the remaining flux exists
as closed magnetic loops).  The fit values of $K_1$ and $m$ quantify
how these processes depend upon the value of $\Upsilon$, for a fixed
rotation rate.

For different rotation rates, the Alfv\'en radius may be modified
further because rotation can act as an additional wind driving
component.  When the stellar rotation is very slow, the rotation has a
negligible affect on the wind driving.  However, for fast rotation,
magnetocentrifugal effects increase the wind velocity.  In order to
quantify this effect, one can think of the wind speed as being
proportional to a rotation-modified speed,
\begin{eqnarray}
\label{eq_vmod}
v_{\rm mod}^2 = K_2^2 v_{\rm esc}^2+ \Omega_*^2 R_*^2
                     = v_{\rm esc}^2 (K_2^2 + 0.5 f^2).
\end{eqnarray}
In this sense, equation (\ref{eq_fitra}) is equivalent to the Alfv\'en
radius being a simple power law in equation (\ref{eq_upsilon}), but
with the factor of $v_{\rm esc}$ being replaced by $v_{\rm mod}$.  The
dimensionless factor $K_2$ determines at what spin rate the stellar
rotation becomes dynamically important for the wind.

\begin{figure}
\epsscale{\psizone}
\plotone{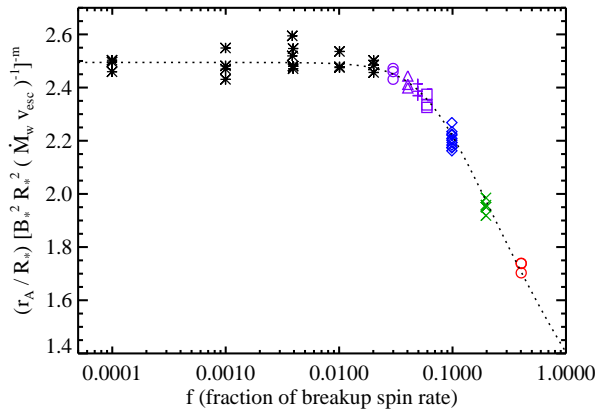}
\caption{Dimensionless Alfv\'en radius times $\Upsilon^{-m}$ versus
  $f$.  This Figure demonstrates the dependence of the Alfv\'en radius
  on the stellar spin rate, for a fixed value of $\Upsilon$.  The
  symbols have the same meaning as Figure \ref{fig_parmspace}, and the
  dotted line shows the fit function (eq.\ [\ref{eq_fitra}] and
  [\ref{eq_fitvals}]).}
\label{fig_Kvf}
\end{figure}

The effect of the stellar spin rate on the Alfv\'en radius (for a
fixed value of $\Upsilon$) is best illustrated by Figure
\ref{fig_Kvf}.  It is clear that when the spin rate is below a few
percent of the breakup rate, the Alfv\'en radius is independent of
spin rate.  Faster rotation decreases the Alfv\'en radius.  For the
fastest spin rates in the parameter study ($f \approx 0.4$), the
Alfv\'en radius is decreased by a factor of approximately 30\%,
relative to the slowly rotating regime.  This corresponds to a factor
of approximately 2 in the torque ($\tau_{\rm w} \propto r_{\rm A}^2$).

\begin{figure}
\epsscale{\psizone}
\plotone{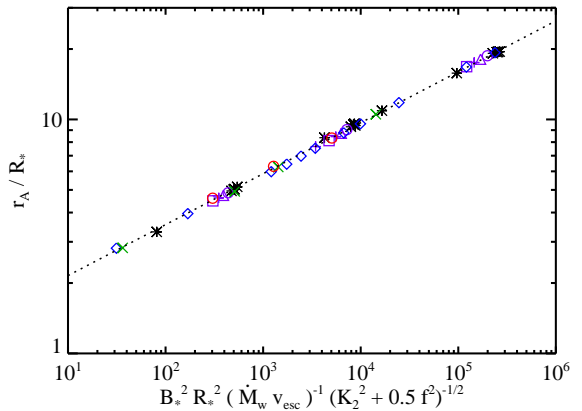}
\caption{Dimensionless Alfv\'en radius versus the quantities in
  square brackets in equation (\ref{eq_fitra}). This Figure demonstrates
  the dependence of the Alfv\'en radius on $\Upsilon$, for a fixed
  stellar spin rate.  The symbols have the same meaning as Figure
  \ref{fig_parmspace}, and the dotted line shows the fit function
  (eq.\ [\ref{eq_fitra}] and [\ref{eq_fitvals}]).}
\label{fig_rafct}
\end{figure}

Figure, \ref{fig_rafct} shows the power-law dependence of the Alfv\'en
radius on $\Upsilon$, for a fixed value of $f$.  It is remarkable how
well equation (\ref{eq_fitra}) fits the dimensionless Alfv\'en radii
determined from all 50 simulations.  As can be seen in Figures
\ref{fig_raplot}--\ref{fig_rafct}, all data points lie within a few
percent (the biggest outlier is off by 4\%) of the fit function.
This reflects the precision of the simulation method in determining
the values of $r_{\rm A}/R_*$ and $\Upsilon$.

By combining equations (\ref{eq_f}), (\ref{eq_upsilon}),
(\ref{eq_rarstar}), and (\ref{eq_fitra}), the torque on the star, due
to the stellar wind, can be written
\begin{eqnarray}
\label{eq_taufit}
\tau_{\rm w} = K_1^2 B_*^{4m} \dot M_{\rm w}^{1-2m} R_*^{4m+2} 
     {{\Omega_*} \over {(K_2^2 v_{\rm esc}^2 + \Omega_*^2 R_*^2)^{m}}}
     \nonumber \\
    = {K_1^2 \over \sqrt{2}}  B_*^{4m} \dot M_{\rm w}^{1-2m} R_*^{4m+1} 
    v_{\rm esc}^{1-2m} {f \over (K_2^2 + 0.5 f^2)^m}
    \nonumber \\
    = {K_1^2 \over (2 G)^m}  B_*^{4m} \dot M_{\rm w}^{1-2m}
    {R_*^{5m+2}  \over M_*^m}
     {\Omega_* \over (K_2^2 + 0.5 f^2)^m},
\end{eqnarray}
where we have listed 3 different equivalent forms for convenience.
This formula is suitable for studies of the evolution of stellar
angular momentum.

\section{Discussion}  \label{sec_discussion}

The torque formulation presented here differs significantly from
analytic prescriptions that have been widely used.  Of particular
note, the preferred model of \citet{Kawaler:1988p1012} results in a
power-law dependence of the Alf\'ven radius on $\Upsilon$, with a
power-law exponent of 0.5.  In the present work, we find this exponent
is approximately 0.22, which results in a significantly different
dependence of the torque on all stellar parameters.  A comparable
exponent of between 0.2 and 0.25 was found in solar-like winds
simulations of \citet{Washimi:1993p3825} and \citet{Pinto:2011p4062},
and \citet{uddoula3ea09} found a similar exponent of 0.25 in
simulations of radiation-driven winds from massive stars.  The reasons
and implications for $m \approx 0.2-0.25$, compared to the analytic
work, was discussed in MP08.
 
We have determined the effect of the stellar rotation rate on the
Alfv\'en radius.  As evident in Figure \ref{fig_Kvf}, there is a slow
magnetic rotator regime, where the rotation rate does not influence
the Alfv\'en radius \citep{Belcher:1976p4310}.  However, for faster
rotation the increased wind acceleration by magnetocentrifugal effects
acts to decrease the Alfv\'en radius.  This effect and transition is
not captured in previous torque formulations.  Note that our
simulation parameter study does not extend fully into the fast
magnetic rotator regime, where the magnetocentrifugal acceleration
completely dominates the thermal driving \citep{Belcher:1976p4310}.
Rather, even for the fastest rotation rates considered here, the
thermal wind driving was not negligible, as appropriate for solar-like
winds.

The torque formulation presented here provides the most physically
realistic and precise calculation to date of solar-like angular
momentum loss, and it is suitable for any studies of the evolution of
angular momentum of sun-like stars.  However, it has a number of
limitations, which point the way for future work.  First of all, the
formulation is not valid in the limit of very weak magnetic fields.
In particular, when the size of the Alfv\'en radius approaches the
stellar radius, the angular momentum transport will begin to be
dominated by other (e.g., viscous) effects, which are not properly
included in our simulations.  To take such effects approximately into
account, we suggest imposing a minimum value of $r_{\rm A} \ge R_*$,
in situations where weak magnetic fields may be considered.

Secondly, the present study fixed the physical parameters that control
the thermal driving of the wind, $\gamma$ and $c_{\rm s}/v_{\rm esc}$
(which are related to the coronal temperature and heating/cooling
physics).  MP08 included a few simulations that had variations in
these wind driving parameters, and we can also compare the torque
formulation presented here with that of \citet{uddoula3ea09}, derived
for radiatively driven winds.  Taken together, it is clear that a
reasonable range in possible wind driving physics can affect the
torque on the star by a factor of $\sim 2$.  It also stands to reason
that the transition from slow magnetic rotator to the regime where
stellar rotation is dynamically important in the wind (i.e., the value
of $K_2$ in eq.\ [\ref{eq_fitra}]) will depend upon the wind driving
physics.  It will be important in future work to be able to reliably
predict how the wind driving physics systematically affect the torque.

Finally, the present study assumed a rotation-axis-aligned, dipolar
magnetic geometry at the stellar surface.  This is justified by the
fact that the largest-scale, global field has the most influence on
the torque, but it is not yet clear how more complex (realistic)
magnetic configurations will change the scalings.  MP08 presented two
simulations with pure quadrupolar magnetic fields.  For the same
surface magnetic field strength $B_*$, the torque for the quadrupole
case was reduced by an order of magnitude compared to the dipole case.
They also noted that the power-law dependence of the Alfv\'en radius
on the strength of the magnetic field may be altered (their 2
simulations suggested an exponent of 0.15, instead of 0.22 for the
dipole).  Similarly, \citet[][]{Pinto:2011p4062} simulated solar winds
with a range of magnetic field complexity, noting a decreased Alfv\'en
radius with increased magnetic complexity and suggesting a different
power-law exponent when the field is significantly non-dipolar.  It
will be important in future work to systematically study how the
torque is affected by complex, realistic magnetic geometries.

\acknowledgments

SPM was supported by the NASA Postdoctoral Program at Ames Research
Center, administered by Oak Ridge Associated Universities through a
contract with NASA, and by the ERC through grant 207430 STARS2
(http://www.stars2.eu).  The National Center for Atmospheric Research
is sponsored by the National Science Foundation.  TPG acknowledges
support from NASA's Origins of Solar Systems program via WBS
811073.02.07.01.89.




\end{document}